# Tunable, synchronized frequency down-conversion in magnetic lattices with defects

Marc Serra-Garcia[1], Miguel Moleron[1] and Chiara Daraio*[2]

[1]Department of Mechanical and Process Engineering, Swiss Federal Institute of Technology (ETH), Zürich, Switzerland

[2]Engineering and Applied Science, California Institute of Technology, Pasadena, California 91125, USA

*email: daraio@caltech.edu

**Abstract -** We study frequency conversion in nonlinear mechanical lattices, focusing on a chain of magnets as a model system. We show that by inserting mass defects at suitable locations, we can introduce localized vibrational modes that nonlinearly couple to extended lattice modes. The nonlinear interaction introduces an energy transfer from the high-frequency localized modes to a low-frequency extended mode. This system is capable of autonomously converting energy between highly tunable input and output frequencies, which need not be related by integer harmonic or subharmonic ratios. It is also capable of obtaining energy from multiple sources at different frequencies with a tunable output phase, due to the defect synchronization provided by the extended mode. Our lattice is a purely mechanical analog of an opto-mechanical system, where the localized modes play the role of the electromagnetic field, and the extended mode plays the role of the mechanical degree of freedom.

**Introduction -** Frequency converting processes have applications in a variety of problems, for example, in obtaining different wavelengths from a fixed-frequency laser[1], harvesting energy from vibration sources[2] and generating entangled photons[3]. Typically, frequency conversion is accomplished through wave mixing[4] (which requires at least two input signals with a prescribed frequency difference), harmonic generation[1] (which produces an output that is a multiple of the input signal) and subharmonic bifurcations[5] (which produce an output that is an integer fraction of the original signal). In addition, these frequency conversion mechanisms prescribe the output's signal phase, which hinders the process of harvesting energy from multiple sources. Combination resonances[6], processes that arise in the presence of multiple nonlinearly-interacting modes, can achieve frequency down-conversion between arbitrary input and output signals not related by a harmonic or subharmonic ratios. The resulting input and output frequencies can be tuned by adjusting the modes' frequencies. Combinational resonances can be found, for example, in vibrating beams[7], membranes and plates[8].

In this paper, we show that nonlinear lattices have the potential to act as frequency-converting devices, due to the combination resonances arising from the nonlinear interaction between the lattice's normal modes. Chains of nonlinearly interacting elements have been studied for decades, beginning in the FPU problem[9,10] . They present a wide variety of phenomena, including solitons[11,12], band-gaps[13], energy trapping[14], breathers[15,16], unidirectional wave propagation[17], negative or extreme stiffness[18], localized modes with tunable profile[19], shocks and rarefaction waves[20]. These phenomena can be used to realize acoustic rectifiers[17], logic gates[21], lenses[22], filters[23], vibration-attenuation[24] and energy harvesting systems[16]. Nonlinear lattices can be implemented using a broad range of materials[25], geometries[26] and interactions[27], allowing to tune the masses, coupling strengths and damping values of the particles, to optimize the performance under the required operating conditions. Because of this tunability, nonlinear metamaterials are a promising candidate for energy converting devices.

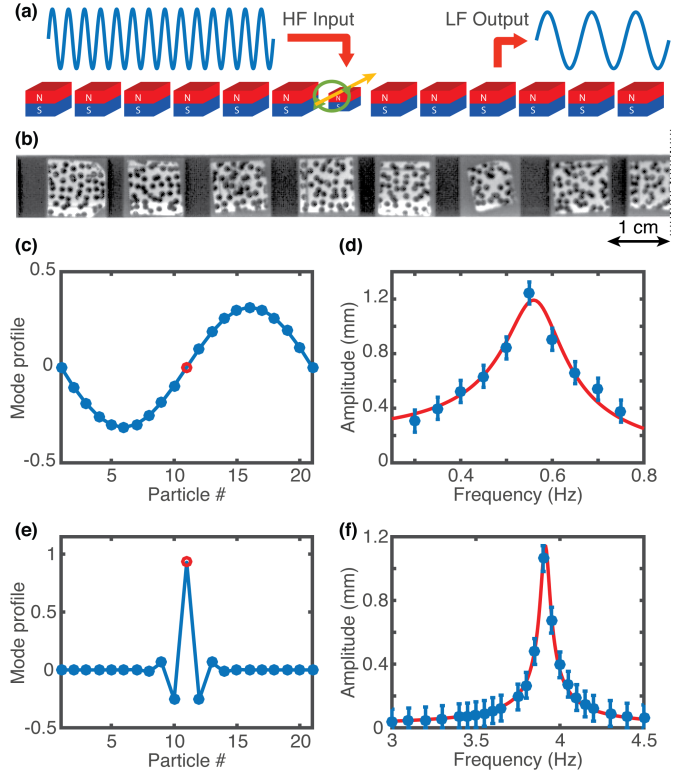

**Figure 1. (Color Online)** *Frequency-converting metamaterial concept.* **(a)** Metamaterial design, consisting of a chain of nonlinearly-interacting magnets. The central particle of the chain is a defect, which has a lower mass. This magnet acts as the high-frequency input to the system. The down-converted energy can be extracted far away from the defect. In our experiments, the defect is driven by a wire carrying an electrical current (yellow arrow). **(b)** Cropped image of the experimental magnet chain, obtained using the same computer vision camera that is also used to track the magnets. Each magnet is enclosed in a 3D printed case, and has a random speckle pattern to facilitate its tracking by digital image correlation. **(c)** Extended mode of vibration. The red hollow circle is the defect particle, while the blue solid dots represent the other particles. **(d)** Experimental frequency response of the extended mode (blue dots) and Lorentzian fit (red solid line). **(e)** Localized mode of vibration. **(d)** Experimental frequency response of the localized mode (blue dots) and Lorentzian fit (red solid line).

**Experimental system –** Our experimental setup consists of a chain of magnets[27] floating on an air table (Fig. 1(a)). Each magnet is embedded in a 3D printed case that adds an additional mass, with different case designs resulting in different particle masses ($m = 0.45g$ for the non-defect particles, $m_{D1} = 0.197g$ for the first defect and $m_{D1} = 0.244g$ for the second defect). The presence of defects introduces localized modes around each defect particle (Fig 1(b,c)). When these modes are excited, the resulting motion is exponentially localized around the defect. In our experimental setup, the defects act as inputs for the frequency-conversion system. We excite them by passing current through a small conductive wire normal to the length of the chain (Fig. 1(a)). The wire is driven harmonically with a signal generator (Agilent 33220A) amplified by an audio amplifier (Topping TP22, class D). An extended mode of vibration (Figs. 1(c) and (d)) interacts with the localized mode to introduce frequency conversion. We monitor the motion of the magnets using a computer vision camera (Point Grey GS3-U3-41C6C-C), with a frame rate between 40 and 200 fps that allows us to resolve all particles' motion. We use the software VIC-2D from Correlated Solutions, to track the particles and determine their trajectory.

**Experimental results for the system with a single defect** – We start by studying a lattice of 21 magnets containing a singe defect ($m_{D1} = 0.197g$) in the middle position ($i=11$). The first and last magnets are fixed. We set the excitation frequency to approximately the sum of the defect's frequency (Fig. 1(f)) and the extended mode's frequency (Fig. 1(d)), with the goal of exciting a combinational resonance ($\omega_E + \omega_L$) between the extended and localized modes [6]. We slowly increase the excitation amplitude until a threshold is reached and self-sustaining oscillations develop far away from the defect, indicating the transfer of energy between the localized mode and an extended mode (Fig. 2(a)). In this regime, the defect motion is modulated by the extended mode (Fig. 2(b)). Due to the exponential localization of the defect's motion, the Fourier transform of a far from the defect particle's displacement (Fig. 2(c)) does not reveal significant motion at the input frequency, and consists almost exclusively of down-converted energy. The frequency conversion efficiency, defined as the energy dissipated in the extended mode in comparison with the energy input into the system $\eta = \frac{<b_L u_L^2>}{\sum_{1 \leq i \leq n}<b_i u_i^2>}$, equals $10.8 \pm 0.9\%$. This efficiency arises from our particular system parameters and is not an absolute limit.

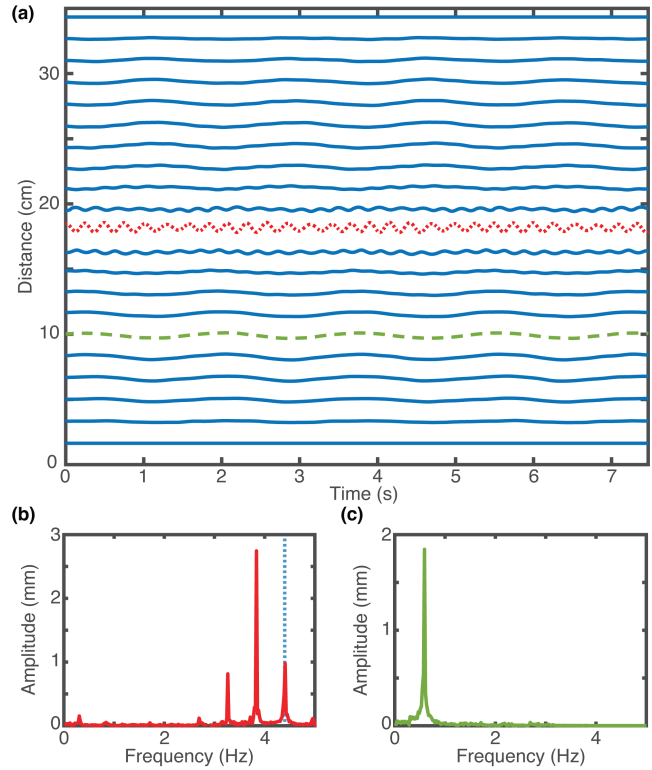

**Figure 2. (Color Online)** *Experimental response of the system under harmonic excitation.* **(a)** Position of the magnets as a function of time. The red dotted line corresponds to the defect magnet, which acts as the input to the frequency-converting system. The green dashed line is taken as the output of the system. **(b)** Fourier transform of the defect magnet's position, which is modulated at the extended mode's frequency. The vertical dotted line represents the excitation frequency. **(c)** Fourier transform of the output magnet's position. This magnet's motion happens primarily at the second extended mode's frequency.

**Theoretical model –** Our theoretical model describes the magnets as point masses. We model the interaction between particles using an empirical power-law model, $F(d) = Ad^B$, with $A = 3.378 \cdot 10^{-12} Nm^{4.316}$ and $B = -4.316$ determined experimentally (See Supplementary Information for the fitted force-displacement curves). This model does not have a straightforward physical justification in terms of the material properties and the geometry of the magnets, but it is chosen because it reproduces the experimental force law with very high precision and low complexity. Using this force-displacement law we can write the equation of motion for the system (the indices in parentheses indicate that no summation is performed over them):

$$m_{(i)}\ddot{x}_{(i)} + b_{(i)}\dot{x}_{(i)} - \sum_{1 \leq j < i} A[d_0(i-j) + x_i - x_j]^B + \sum_{i < j \leq n} A[d_0(j-i) + x_j - x_i]^B = F_i(t) \quad (1)$$

Where $m_{(i)}$ and $b_{(i)}$ are the mass and damping coefficient of the $i$-th particle, $A$ and $B$ are the magnetic force law parameters, $F_i(t)$ is the external driving force acting on the $i$-th particle (which may be zero if the particle is not externally driven), and $d_0$ is the distance between magnets at rest. When performing the reduced-order analysis, we will assume that $d_0$ is the same for all magnets. This is an approximation, since magnets that are not in the center of the lattice are subject to asymmetric long-range forces. However, we have found this approximation to yield acceptable results. We emphasize that our theoretical model is not limited to nearest-neighbor interactions and takes into account the magnetic force between all pairs of magnets. All numerical integration in this paper is done using a 4[th] order Runge-Kutta algorithm with a time step of $1\ ms$.

**Reduced modal description and frequency conversion mechanism –** The mechanism responsible for the frequency conversion in our lattice becomes much clearer when we look at the evolution of the system in terms of the normal modes of the linearized system. We can obtain this description by approximating the force-displacement relation by a second order Taylor series. When we do this approximation, the system becomes:

$$M_{ij}\ddot{x}_j + B_{ij}\dot{x}_j + K_{ij}x_j + \Gamma_{ijk}x_jx_k = F_i(t) \quad (2)$$

Here, the indices $j$ and $k$ are summed over all degrees of freedom, $M$, $B$ denote the mass and damping matrices defined conventionally, and the terms $K$ and $\Gamma$ are obtained by Taylor expansion of the force law:

$$K_{ij} = \frac{d}{dx_j}\left(\sum_{1 \leq m < i} A[d_0(i-m) + x_i - x_m]^B + \sum_{i < m \leq n} A[d_0(m-i) + x_m - x_i]^B\right) \quad (3)$$

$$\Gamma_{ijk} = \frac{1}{2}\frac{d^2}{dx_j dx_k}\left(\sum_{1 \leq m < i} A[d_0(i-m) + x_i - x_m]^B + \sum_{i < m \leq n} A[d_0(m-i) + x_m - x_i]^B\right) \quad (4)$$

Since $M$ is symmetric and positive-definite and $K$ symmetric, we can find an invertible matrix $P$ such that $P^T M P$ and $P^T K P$ are both diagonal. For simplicity, we assume that damping is proportional to the mass matrix and therefore also diagonalizable. In this basis, the equations of motion become:

$$P_{im}M_{ij}P_{jn}\ddot{u}_n + P_{im}B_{ij}P_{jn}\dot{u}_n + P_{im}K_{ij}P_{jn}u_n + \Gamma_{ijk}P_{im}P_{jn}P_{kl}u_n u_l = F_m(t) \quad (5)$$

The diagonalized system in Eq. (5) does not provide any significant numerical advantage, since $\Gamma_{ijk}$ is highly non-local in the modal basis (i.e., modes far apart interact as strongly as nearby modes). However, we can see the motivation for this approach if we look at the experimental results in the modal basis (Fig. 3(a)). In this basis, most of the motion occurs in the second extended mode and in the localized mode. In fact, these modes hold around 90% of the system's energy (Fig. 3(b)). Therefore, we can restrict our description to these two modes without incurring a significant error. This reduced-order description is:

$$m_L \ddot{u}_L + b_L \dot{u}_L + (k_L - 2\gamma u_E)x_L = F_I \cos(2\pi f_I t) \quad (6)$$

$$m_E \ddot{u}_E + b_E \dot{u}_E + k_E u_E - \gamma u_L^2 = 0 \quad (7)$$

In this description, $u_L$ and $u_E$ are the displacements of the localized and extended modes respectively, $F_I$ is the input force, $f_I$ is the input frequency, $m_L$, $b_L$ and $k_L$ are the effective mass, damping and stiffness of the localized defect mode, $m_E$, $b_E$ and $k_E$ are the effective mass, damping and stiffness of the extended mode. The term $\gamma = \Gamma_{ijk}P_{iE}P_{jL}P_{kL} = \Gamma_{ijk}P_{iL}P_{jE}P_{kL} = \Gamma_{ijk}P_{iL}P_{jL}P_{kE}$ denotes the quadratic interaction between modes. This term can be determined by performing a Taylor expansion of the interaction force, or by analyzing the frequency response of the local and extended modes (See Supplementary Information). We note that the reduced equations of motion present an asymmetry: There is no term proportional to $x_E^2$ in Eq. (6) and there is no term proportional to $x_E x_L$ in Eq. (7). These terms do not appear in our lattice due to the location of the defect, but they are not generally zero (See Supplementary Information for a study on the relation between nonlinear terms and defect location). The interaction between modes can be understood in the following way: Due to nonlinearity, the vibration of the defect mode pushes against its neighbors, in a way that is analogous to thermal expansion of a crystal[18] or the optical pressure in an opto-mechanical system (Fig. 3(c)). For small amplitudes, this expansion is proportional to the square of the vibration amplitude, resulting in the term $\gamma x_L^2$ in the extended mode equation. In addition, the motion of the extended mode modulates the distance between the defect particle and its neighbors (Fig. 3(d)). This affects the localized mode's effective stiffness and introduces the parametric term $\Delta k_L = \gamma x_E$, analogous to the modulation of the optical cavity wavelength in an opto-mechanical system. This type of interaction appears in a variety of systems, such as phonon modes in superconductors[28], and can result in stochastic heat engine operation[29]. This reduced-order model can reproduce the experimentally-observed behavior (Figs. 3(e) and (f)) with remarkable accuracy. We highlight that the only fitting parameter used is the excitation amplitude. The particle mass, mode quality factor and inter-particle force law have all been measured experimentally.

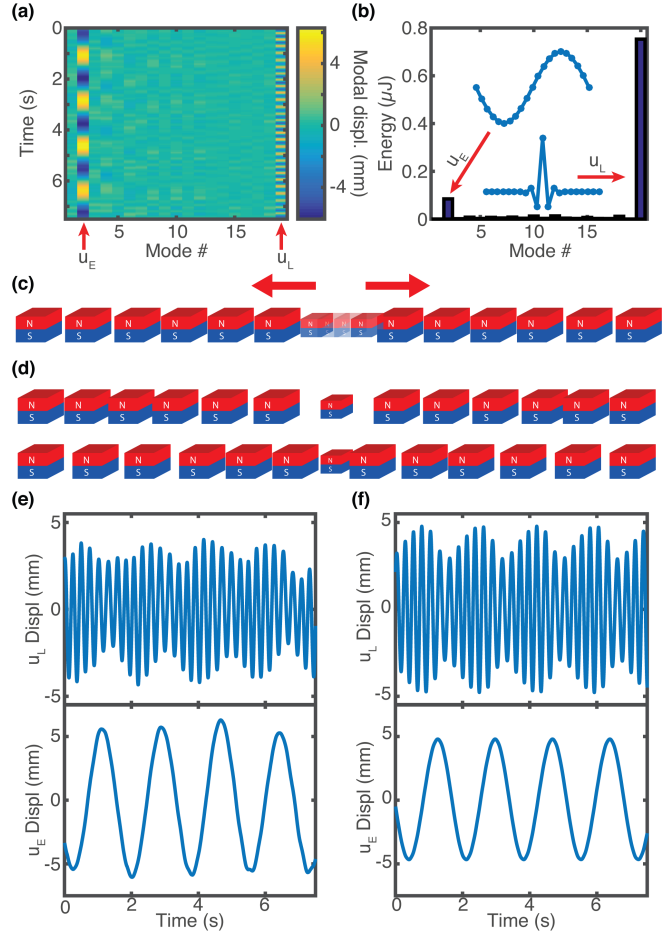

**Figure 3. (Color Online)** *Reduced-order description of the frequency conversion process.* **(a)** Projection of the experimental time evolution (Fig. 2(a)) in the linear modal basis. **(b)** Average energy as a function of the mode number. The system's energy is highly concentrated in the second extended mode and the localized defect mode. **(c)** Dynamic expansion of the defect mode. When the defect vibrates, the nonlinear magnetic interaction results in a non-zero average force acting on the defect's neighbors. **(d)** The motion of the second extended mode modulates the distance between the defect particle and its neighbors, dynamically tuning the defect mode frequency. **(e)** Detail of the extended mode and localized mode evolution, measured experimentally. **(f)** Theoretical prediction for the extended and localized mode evolution, obtained from a reduced-order model considering only two modes (Eq. (6) and (7)). The numerical simulation in panel **f** corresponds to a system with $m_E = 0.45\ g$, $m_L = 0.232\ g$, $f_E = 0.5664\ Hz$, $f_L = 3.913\ Hz$, $f_I = 4.38\ Hz$, $F_I = 45\ \mu N$, $Q_E = 4.518$, $Q_L = 66.62$ and $\gamma = 1.801\ N/m^2$, where $k_X = m_X(2\pi f_X)^2$ and $b_X = m_X 2\pi f_X / Q_X$.

The two-mode system, described in Eq. 6 and Eq. 7, is a purely-mechanical analog of an opto-mechanical system[30-32]. The extended mode plays the role of the low-frequency mechanical motion, while the localized mode plays the role of the high-frequency electromagnetic field. The term $\gamma x_L^2$ acting on the extended mode plays the role of the optical pressure, while the term $2\gamma x_E x_L$ acting on the localized mode reproduces the modulation of the Fabry-Perot resonance by the mechanical degree of freedom in an opto-mechanical system. This analogy can be made explicit by expressing the motion of the defect mode as $u_L(t) = (1/2)[a(t)e^{i\omega t} + a^*(t)e^{-i\omega t}]$ and assuming that $a(t)$ changes slowly and that $1/Q_L \ll 1$. With these assumptions, we arrive at the following equation (a detailed derivation and comparison with the full model are provided in the Supplementary Information):

$$m\ddot{u}_E + b_E\dot{u}_E + k_E u_E = \gamma\frac{|a|^2}{2} \tag{8}$$

$$\dot{a} + a\left[\frac{\omega_0}{2Q_L} - i\Delta(u_E)\right] = F_I \tag{9}$$

Here, $\omega_0$ is the natural frequency of the localized mode and the detuning $\Delta(u_E)$ is the difference between the localized mode's natural frequency and the defect's excitation frequency, as a function of the extended mode's position. All other parameters have the same meaning than they did in Eq. (6) and Eq. (7). While being an approximation, this form has numerical advantages by not containing rapidly changing components at the frequency of the localized mode, and not requiring the evaluation of trigonometric functions for the excitation. Besides numerical reasons, the description provided in Eq. (8) and Eq. (9) is identical to the model of an opto-mechanical system[30,32,33], for which there is extensive analytical literature[32,34]. This analogy provides a lucid interpretation of the frequency converting mechanism, whereby the self-sustaining oscillations of the extended more are the result of a feedback mechanism between the extended mode's motion and the localized mode amplitude. In this picture, the localized mode amplitude $a$ depends on the extended mode displacement through the term $\Delta(u_E)$. Equation (9) imposes a retardation between $a$ and $u_E$ and, as a consequence, the term $\gamma|a|^2$ has a quadrature component (shifted 90 degrees from $u_E(t)$) that results in negative damping[31]. When this negative damping exceeds the value of $b_E$, the system develops self-sustaining oscillations, which saturate at a finite value due to non-linearity[31].

**Multiple-defect synchronized frequency conversion –** Systems containing multiple defects can present synchronized frequency conversion, where the motion of multiple defects is determined by the same extended mode, thereby synchronizing the defect's modulation envelopes and resulting in the conversion of energy from multiple input frequencies to a single output frequency. The use of an extended mode to synchronize multiple resonant elements appears in the context of Josephson junction arrays[35], and here we use it to extract energy from multiple sources of mechanical vibrations. Our experimental system consists of 20 magnets, with defect particles in positions 7 ($0.244g$) and 14 ($0.197g$). The initial and final particles are fixed. As in the case with a single defect, we set an excitation frequency for each defect equal to the defect's frequency plus an extended mode's frequency. This time we use the third extended mode instead of the second, because it presents two regions of maximum strain at the two defect's positons. As we did in the single defect case, we increase both defect's excitation amplitudes simultaneously, until we observe self-sustaining oscillations far from the defect. Figure 4a shows the trajectories of the magnets in the self-sustaining regime. We calculate the energy transfer between both defects and the extended mode, by utilizing the empirical force-displacement relation and the defect's trajectories, and we observe that both defects are contributing energy to the extended mode with a power ($P = <F\dot{x}_E> = <\gamma x_L^2 \dot{x}_E>$) of $16.9 \pm 1.5\ nW$ and $25.8 \pm 4.0\ nW$ respectively, indicating successful extraction of energy from multiple sources. The frequency conversion efficiency is $20.5 \pm 10.4\%$. As in the previous case, the motion of the defects presents sidebands indicating a modulation by the extended mode (Fig. 4(b)). Far from the defect, the motion takes place exclusively at the extended mode's frequency, as required for successful frequency conversion operation.

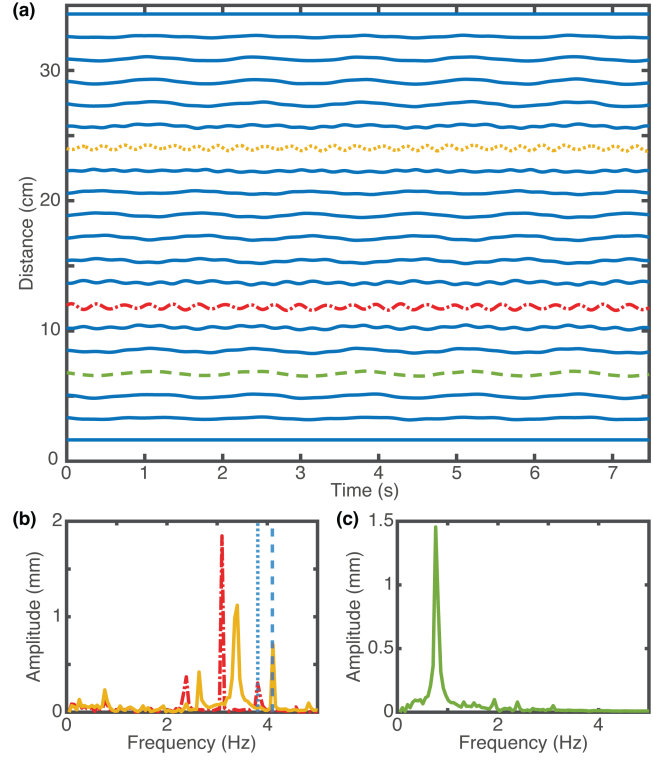

**Figure 4. (Color Online)** *Synchronized frequency conversion.* **(a)** Position of the magnets as a function of time. The yellow dotted line (particle 7) and the red dotted-dashed line (particle 14) are defect magnets that act as the high-frequency inputs of the system. The green dashed line is the low frequency output. **(b)** Fourier transform of the defects' positions, which are modulated at the extended mode's frequency. The vertical dotted line represents the excitation frequency. **(c)** Fourier transform of the output magnet's position. This magnet's motion happens primarily at the third extended mode's frequency.

As in the case with a single defect, expressing the magnet's trajectories in terms of the lattice's linear normal modes reveals that the motion (Fig. 5(a)) and the energy (Fig. 5(b)) are primarily concentrated in an extended mode (Fig. 5(c), left) and in the two localized modes, the profiles of which are depicted in Fig. 5(c). This energy concentration allows us to formulate a reduced-order description following the same procedure as in the system with a single defect. The resulting system of equations has the form:

$$m_{L1}\ddot{x}_{L1} + b_{L1}\dot{x}_{L1} + (k_{L1} - 2\gamma_1 x_E)x_{L1} = F_{I1}\cos(2\pi f_{I2} t) \quad (10)$$

$$m_{L2}\ddot{x}_{L2} + b_{L2}\dot{x}_{L2} + (k_{L2} - 2\gamma_2 x_E)x_{L2} = F_{I2}\cos(2\pi f_{I2} t) \quad (11)$$

$$m_E\ddot{x}_E + b_E\dot{x}_E + k_E x_E - \gamma_1 x_{L1}^2 - \gamma_2 x_{L2}^2 = 0 \quad (12)$$

The model in Eqs. (10)-(12) is capable of qualitatively predicting the evolution of the modes (Fig. 5(d) and 5(e)), but under-estimates the output amplitude relative to the experiments. We attribute this difference to uncertainty in the system's resonance frequencies and quality factors. This is suggested by the difference between theory and experiment in the extended mode's frequencies (See Supplementary Information) and in the phase of the localized mode's vibration.

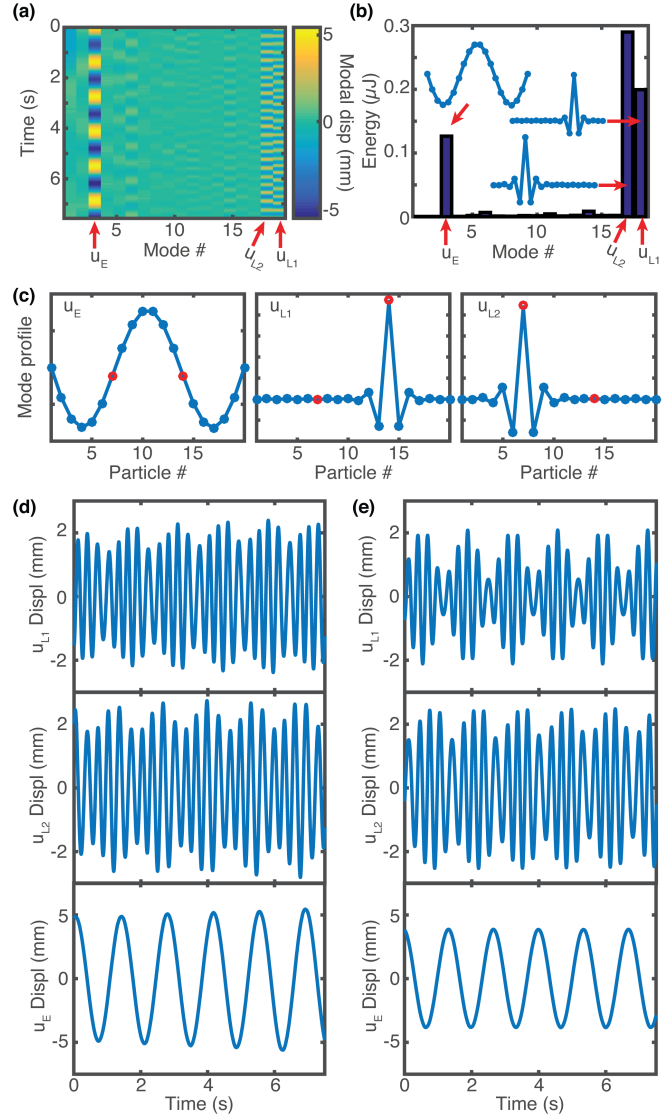

**Figure 5. (Color Online)** *Reduced-order description of the synchronized frequency conversion.* **(a)** Time evolution of the magnets in terms of the linear eigenmode basis. **(b)** Energy distribution in each normal mode. The energy is concentrated in the third extended mode and in the two localized defect modes. **(c)** Mode profiles of the three most relevant eigenmodes. **(d)** Experimental time evolution of the third extended mode $u_E$ and the two localized modes $u_{L1}$ and $u_{L2}$ as a function of time. **(e)** Theoretical prediction for the time evolution of the eigenmodes. The theoretical predictions have been obtained using a 3-DOF reduced order model. The numerical parameters used in panel **(e)** are: $m_E = 0.45\ g$, $m_{L1} = 0.2318\ g$, $m_{L2} = 0.2915\ g$, $f_E = 0.7494\ Hz$, $f_{L1} = 3.404\ Hz$, $f_{L2} = 3.063\ Hz$, $f_{I1} = 4.1\ Hz$, $f_{I2} = 3.81\ Hz$, $F_{I1} = F_{I2} = 42\ \mu N$, $Q_E = 12.27$, $Q_{L1} = 39.3$, $Q_{L2} = 60.27$, $\gamma_1 = -2.4293\ N/m^2$, $\gamma_2 = 2.5761\ N/m^2$.

**Output phase tunability –** In our lattice, the output signal's phase is not prescribed by the inputs and can be dynamically tuned while the system is operating. This offers an opportunity to synchronize multiple devices, create passive and tunable phased arrays or transfer information by modulating the output signal's phase. We theoretically demonstrate this output phase tunability in Fig. 6(a)-(c). The phase modulation is accomplished by perturbing the last particle following a Gaussian profile given by $x_n(t) = A_P e^{-\frac{(t-t_0)^2}{2\sigma^2}}$ (Fig. 6(a)), where $A_P$ denotes the maximum perturbation amplitude, $t_0$ is the moment where the perturbation is applied and $\sigma$ represents the width of the perturbation. We choose a Gaussian profile because it is highly localized in both time and frequency domains. Applying this perturbation results in a change in the output signal phase, that persists long after the perturbation has vanished. Figure 6(b) shows the extended mode's displacement 1790 seconds after a perturbation, for different perturbation amplitudes. In this calculation, the perturbation width $\sigma = 30s$ is much smaller than the wait time, ensuring that no effect remains by the time the results are obtained. Two remarkable observations shall be made regarding the phase tunability shown in Fig. 6(c): Firstly, this tunability covers the whole range ($0° - 360°$). Secondly, the perturbation-induced phase shift persists for a period of time that is much longer than any of the system's time constants, since the phase does not change significantly if we wait an additional $1000s$ until $t = 2790s$. In the experimental system, this phase stability would be limited by the presence of external noise sources and Brownian motion.

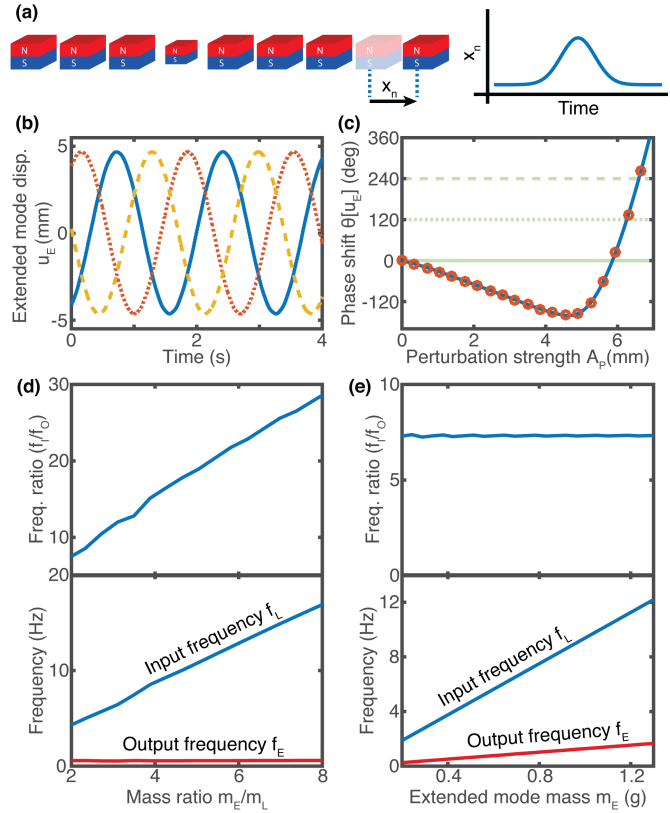

**Figure 6. (Color Online)** *Theoretical investigation of phase and frequency tunability.* **(a)** Phase tuning scheme. The output signal's phase is tuned by moving the last particle ($x_n$) following a Gaussian profile. **(b)** Extended mode signal 2790 seconds after the phase-shifting perturbation has been effected. The lines correspond to perturbations with $A_0$ equal to 0 mm (blue, solid), 6.2562 mm (red, dotted) and 6.5917 mm (yellow, dashed). **(c)** Output phase as a function of the maximum displacement of the phase-adjusting perturbation. The blue solid line is measured 1790 seconds after the perturbation, while the circles are measured 1000 seconds after the first measurement, 2790 seconds after the perturbation's peak. Panels **b** and **c** have been obtained by integrating the full equation of motion (Eq. (1)) with $d_0 = 16.3\ mm$, $m_{i,i \neq 11} = 0.45\ g$, $m_{11} = 0.197\ g$, $b_{i,i \neq 11} = 306.83\ \mu Ns/m$, $b_{11} = 42.62\ \mu Ns/m$, $F_{i,i \neq 11} = 0N$, $F_{11} = F_I \sin 2\pi f_i t$, $F_I = 48.45\ \mu N$ and $f_I = 4.38\ Hz$. The force-law parameters are as described in the theoretical model section. **(d)** Frequency down-conversion ratio (top) and input and output frequencies (bottom) as a function of the mass ratio between the defect and extended modes. These plots have been obtained by keeping the extended mode's mass constant and modifying the defect's mass. **(e)** Frequency down-conversion ratio (top) and input and output frequencies (bottom) as a function of the extended mode mass, while keeping

the modal mass ratio $m_E/m_L$ constant. In this section, all parameters except the masses are identical to those in Fig. **3(f)**.

**Tunability –** A remarkable feature of our frequency-converting system is the possibility of tuning the input and output frequencies over a broad range, both during the design phase and dynamically once the system has already been built. Figure (6) theoretically explores the relationship between the input and output frequencies and the modal masses. We first explore the effect of the mass ratio by altering the mass of the defect mode without altering that of the extended mode (Fig. 6(a)). This results in a change in the optimal input frequency without a significant effect on the output frequency. We then proceed to alter the masses of the localized and extended mode simultaneously (Fig. 6(b)). This affects both input and output frequencies, while maintaining the down-conversion ratio constant. In the conversion ratio calculation, we identify the optimal input frequency by sweeping the input between the resonance frequency of the localized mode and the resonance frequency of the localized mode plus twice the resonance frequency of the extended mode, and finding the input frequency that results in the highest energy transfer. In addition to the particle's mass, there are many unexplored avenues for tuning the frequency conversion ratio. Examples include the static compression applied on the chain, the magnets' strength and geometry and the application of an external magnetic fields[36]. In addition, modern 3D printed materials allow us to engineer nonlinear inter-particle interactions[37] beyond these offered by magnetic systems.

**Conclusions and outlook –** We have demonstrated that lattices composed of magnetically interacting particles with defects are capable of converting energy from high frequencies to lower frequencies, which need not be linked by harmonic/subharmonic relations. This is possible through the nonlinear coupling between extended and localized normal modes. Such frequency-converting lattice, analogous to opto-mechanical systems, is highly tunable in both frequency and phase, and can extract energy from multiple signals at different frequencies to generate a single-component output. This work may motivate the design of innovative nonlinear metamaterials and devices with tunable energy conversion capabilities.

# Supplementary Information: Tunable, synchronized frequency down-conversion in magnetic lattices with defects

Marc Serra-Garcia[1], Miguel Moleron[1] and Chiara Daraio*[2]

[1]Department of Mechanical and Process Engineering, Swiss Federal Institute of Technology (ETH), Zürich, Switzerland

[2]Engineering and Applied Science, California Institute of Technology, Pasadena, California 91125, USA

*email: daraio@caltech.edu

## Force-displacement relation

We experimentally determine the parameters for the interaction force by measuring the force-displacement relation and fitting it using a power law function of the form $F = Ax^B$. The measurements and fitted curve are plotted in Fig. S1. Through this procedure, we obtain values of $A = 3.378 \pm 0.751 \, pNm^{4.316}$ and $B = -4.316 \pm 0.0460$.

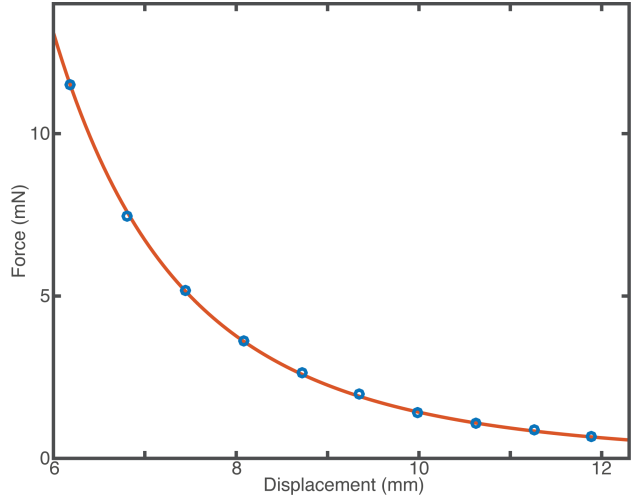

**FIG S1:** Magnetic *force-displacement relation.* The blue circles are the experimental measurements and the red line represents the power-law fit.

## Derivation of the optomechanical equations of motion

Here we show that the equations governing the reduced-order dynamics of the frequency converting lattice (Eq. 6-7 in the main paper) can be approximated by those describing an optomechanical system, under realistic assumptions. The equations that we want to approximate are:

$$m_L \ddot{u}_L + b_L \dot{u}_L + (k_L - 2\gamma u_E)u_L = F_I \cos(\omega t) \qquad \text{Eq. S1}$$

$$m_E \ddot{u}_E + b_E \dot{u}_E + k_E u_E - \gamma u_L^2 = 0 \qquad \text{Eq. S2}$$

We start our approximation by expressing the displacement of the localized mode as a harmonic function, with a slowly-changing amplitude and phase given by the complex function $a(t)$:

$$u_L = \frac{1}{2}\left[a(t)e^{i\omega t} + \bar{a}e^{-i\omega t}\right] \qquad \text{Eq. S3}$$

This allows us to rewrite Eq. S1 as:

$$m_L \frac{d^2}{dt^2} \frac{1}{2}\left[a(t)e^{i\omega t} + \bar{a}(t)e^{-i\omega t}\right] + \frac{m_L \omega_{0L}}{Q_L} \frac{d}{dt} \frac{1}{2}\left[a(t)e^{i\omega t} + \bar{a}(t)e^{-i\omega t}\right] + \qquad \text{Eq. S4}$$

$$(k_L - 2\gamma u_E)\tfrac{1}{2}[a(t)e^{i\omega t} + \bar{a}(t)e^{-i\omega t}] = \tfrac{1}{2}[F_0 e^{i\omega t} + \bar{F}_0 e^{-i\omega t}]$$

For equation S4 to hold, terms multiplied by $e^{i\omega t}$ must be identical on both sides of the equation (Taking terms multiplied by $e^{-i\omega t}$ would yield an identical condition. This identity results in:

$$m_L \frac{d^2}{dt^2}[a(t)e^{i\omega t}] + \frac{m_L \omega_{0L}}{Q_L}\frac{d}{dt}[a(t)e^{i\omega t}] + (k_L - 2\gamma u_E)a(t)e^{i\omega t} = F_0 e^{i\omega t} \quad \text{Eq. S5}$$

By evaluating the derivative and both sides by $e^{i\omega t}$ we obtain:

$$m_L[-\omega^2 a(t) + 2i\omega \dot{a}(t) + \ddot{a}(t)] + \frac{m_L \omega_{0L}}{Q_L}[i\omega a(t) + \dot{a}(t)] + (k_L - 2\gamma x_E)a(t) = F_0 \quad \text{Eq. S6}$$

Now we make our first assumption: That the localized mode amplitude and phase are slowly changing. This allows us to neglect the second derivative of $a$ in Eq. S6, resulting in:

$$m_L[-\omega^2 a(t) + 2i\omega \dot{a}(t)] + \frac{m_L \omega_{0L}}{Q_L}[i\omega a(t) + \dot{a}(t)] + (k_L - 2\gamma x_E)a(t) = F_I \quad \text{Eq. S7}$$

Which can be regrouped as:

$$\left[\frac{\omega_{0L}}{Q_L} + 2i\omega\right]\dot{a}(t) + \left[\frac{k_L - 2\gamma u_E}{m_L} - \omega^2 + \frac{\omega_{0L}}{Q_L}i\omega\right]a(t) = \frac{F_I}{m_L} \quad \text{Eq. S8}$$

Since $Q_L$ is a large value (40-60 in our experimental setup) and $\omega_{0L}$ is of the same order than $\omega$, we can neglect the contribution of $\omega_{0L}/Q_L$ in the right hand side. This results in:

$$\dot{a}(t) + \left[\frac{-i}{2\omega}\left(\frac{k_L - 2\gamma u_E}{m_L} - \omega^2\right) + \frac{\omega_{0L}}{2Q_L}\right]a(t) = \frac{F_I}{2\omega m_L} \quad \text{Eq. S9}$$

We note that the term $(k_L - 2\gamma x_E)/m_L$ is the square of the localized mode's angular frequency, as a function of the displacement of the extended mode. We call this term $\left(\omega_{0L}(x_E)\right)^2$ to distinguish it from the natural frequency of the localized mode when the extended mode is at rest ($x_E = 0$), which we termed $\omega_{0L}$. By defining the detuning $\Delta(x_E) = \omega_{0L}(x_E) - \omega$ as the difference between the natural frequency $\omega_{0L}(x_E)$ and the excitation frequency, we obtain:

$$\dot{a}(t) + \left[\frac{-i}{2(\omega_{0L}(u_E) - \Delta(u_E))}\left((\omega_{0L}(u_E))^2 - (\omega_{0L}(u_E) - \Delta(u_E))^2\right) + \frac{\omega_{0L}}{2Q_L}\right]a(t) = \frac{F_I}{2\omega m_L} \quad \text{Eq. S10}$$

Expanding the squares in the left hand side of Equation 10, we obtain:

$$\dot{a}(t) + \left[\frac{i(\Delta(u_E) - 2\omega_{0L}(u_E))\Delta(u_E)}{2(\omega_{0L}(u_E) - \Delta(u_E))} + \frac{\omega_{0L}}{2Q_L}\right]a(t) = \frac{F_I}{2\omega m_L} \quad \text{Eq. S11}$$

Since $\Delta \ll \omega_{0L}(x_E)$ we can neglect the additive term $\Delta(x_E)$ in the numerator and denominator:

$$\dot{a}(t) + \left[\frac{i(-2\omega_{0L}(u_E))\Delta(u_E)}{2(\omega_{0L}(u_E))} + \frac{\omega_{0L}}{2Q_L}\right]a(t) = \frac{F_I}{2\omega m_L} \quad \text{Eq. S12}$$

Which can be simplified to the equation for a classical optomechancial system:

$$\dot{a}(t) + \left[\frac{\omega_{0L}}{2Q_L} - i\Delta(u_E)\right]a(t) = \frac{F_I}{2\omega m_L} \quad \text{Eq. S13}$$

We now examine the equation for the extended mode, by replacing $u_L(t)$ (Eq. S3) into Eq. S2.:

$$m_E \ddot{u}_E + b_E \dot{u}_E + k_E u_E - \frac{\gamma}{4}\left(a^2 e^{i2\omega t} + \bar{a} e^{-i2\omega t} + 2a\bar{a}\right) = 0 \quad \text{Eq. S14}$$

By neglecting the rapidly varying degrees of freedom at 2ω, we arrive at the equation:

$$m_E \ddot{u}_E + b_E \dot{u}_E + k_E u_E - \frac{\gamma}{2}|a|^2 = 0 \quad \text{Eq. S15}$$

Equations S13 and S15 correspond to the optomechanical model presented in the main paper.

# Tuning the nonlinear parameters by modifying the defect's location

We can tune the nonlinear parameters in our reduced-order model by changing the defect location. The most general equation of motion for the system, truncated to contain only second-order terms, is given by:

$$m_L \ddot{u}_L + b_L \dot{u}_L + k_L u_L - \beta_L u_L^2 - 2\gamma_E u_L u_E - \gamma_L u_E^2 = F_I \cos(\omega t) \quad \text{Eq. S16}$$

$$m_E \ddot{u}_E + b_E \dot{u}_E + k_E u_E - \beta_E u_E^2 - 2\gamma_L u_L u_E - \gamma_E u_L^2 = 0 \quad \text{Eq. S17}$$

Figure S2 presents the nonlinear parameters as a function of the defect location. The selected defect locations maximize the $\gamma_E$ coupling, while ensuring that all other nonlinear parameters are small. The point of maximal $\gamma_E$ corresponds to the region where the mode's strain $\epsilon_i = x_{i+1} - x_i$ is maximal, resulting in the highest change in the defect-neighbor distance during the motion of the extended mode.

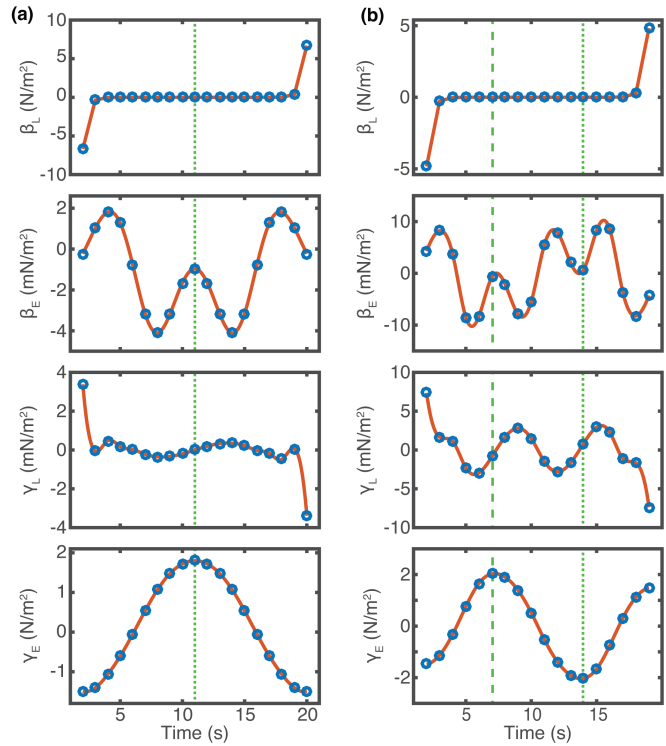

**FIG S2:** *Nonlinear parameters as a function of the defect location.* **(a)** Here, the extended mode is the second extended mode of the lattice, corresponding to the single-defect system in the main paper. **(b)** The extended mode is the third extended mode of the lattice, corresponding to the two-defects system in the main paper. In both panels, the dotted line represents the experimental light defect location. In panel **b**, the dashed line represents the heavy defect location.

## Comparison between full, reduced and optomechanical system

Here we present a comparison between the system's evolution predicted by the full model (Eq. XX in the main paper), the two-mode reduced order model (Eq. 1 in the main paper) and the optomechanical model (Eq. 6 and Eq. 7 in the main paper).

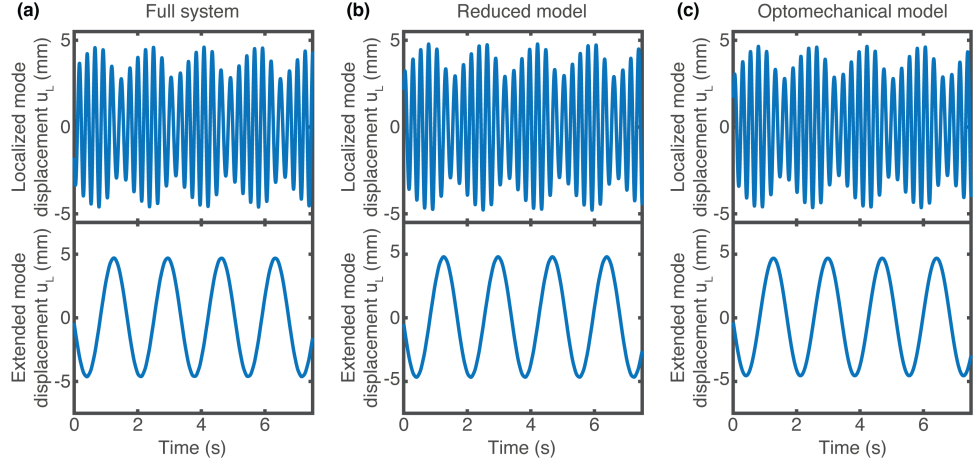

**FIG S3:** *Comparison of full and reduced models.* **a** Time evolution of the localized and extended modes calculated using the full system simulation. The modal description has been obtained by projecting the trajectories into the modal basis. **b** Modal time evolution calculated using the two-mode reduced order model. **c** Modal time evolution calculated using the optomechanical model. The $t = 0$ point has been selected independently in each simulation, in order to present a consistent phase.

We observe that the three models produce similar predictions. This allows us to conclude that a reduced-order modelling approach provides a good approximation, and that our nonlinear lattice accurately mimics the dynamics of an optomechanical system.

## Determination of the natural frequencies in the two-defect system

Here we discuss the determination of the resonance frequencies and quality factors for the third extended mode and the two localized modes, used in the section *Multiple-defect syncronized frequency conversion* of the main paper. The frequency is determined by exciting the modes using a variable frequency signal. We monitor each particle's motion and project it into the theoretically-predicted modal basis. The amplitude is determined by calculating the RMS value of the modal coordinate after subtracting the average. We then fit the frequency response using a Lorentzian function to obtain the mode's frequency and quality factor.

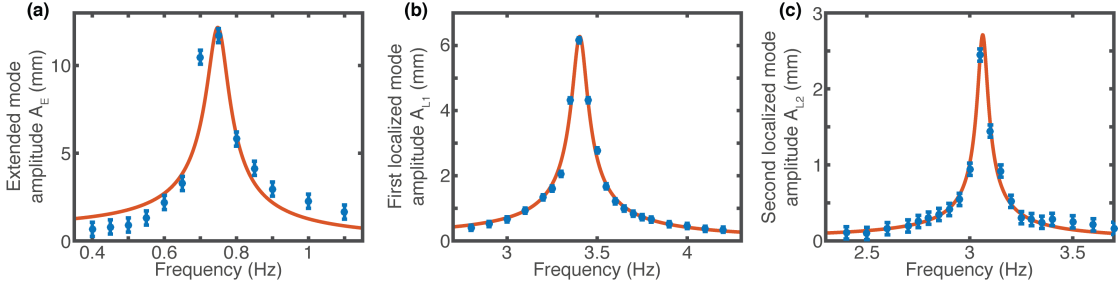

**FIG S4:** *Fitting of the two-defect system parameters.* **a** Frequency response of the third extended mode. **b** Frequency response of the first localized mode (Centered around the defect with mass $m_{D1} = 0.197\ g$. **c** Frequency response of the second localized mode (Centered around the defect with mass $m_{D1} = 0.244\ g$

| | | |
|---|---|---|
| Extended mode | Frequency | $0.7494 \pm 0.0197\ Hz$ |
| | Quality factor | $12.27 \pm 6.24$ |
| First Localized mode | Frequency | $3.404 \pm 0.004\ Hz$ |
| | Quality factor | $39.30 \pm 3.35$ |
| Second localized mode | Frequency | $3.063 \pm 0.004\ Hz$ |
| | Quality factor | $60.27 \pm 10.34$ |

**Table 1:** *Two-defect system model parameters.*

# Determination of the nonlinear constant from the frequency response

In all of our paper's simulations, the nonlinear parameter $\gamma$ has been determined by performing a Taylor expansion of the magnetic force-displacement relation presented in Fig. S1. In some circumstances (For example, in microscopic systems) it may not be possible to accurately measure the interaction potential. Here we calculate the nonlinear parameter $\gamma$ from the frequency response curves (Fig. S5a), by simultaneously monitoring the displacement of the extended mode (Fig. S5b) during the frequency response characterization.

The equation of motion for the extended mode is given by:

$$m_E \ddot{u}_E + b_E \dot{u}_E + k_E u_E - \gamma u_L^2 = 0 \qquad \text{Eq. S18}$$

For excitation amplitudes below the self-oscillation threshold, $u_L$ follows a harmonic motion with constant amplitude. Under these conditions, $u_E$ cannot follow the rapid changes of $u_L^2$ and reacts only to its average value. Since the displacement of $u_E$ during the frequency response measurement is quasistatic, we can neglect the terms $m_E \ddot{u}_E$ and $b_E \dot{u}_E$. This results in the equation:

$$< u_E >= \left(\frac{\gamma}{k_E}\right) < u_L^2 >$$

Eq. S19

Figure S5c presents the extended mode displacement as a function of the localized mode amplitude. By fitting this relation using a quadratic polynomial, we obtain a nonlinear coefficient $\gamma = 1.79 \pm 0.56\ N/m^2$ which compares extremely well with the value $\gamma = 1.81 \pm 0.42\ N/m^2$ obtained from the experimental force-displacement relation.

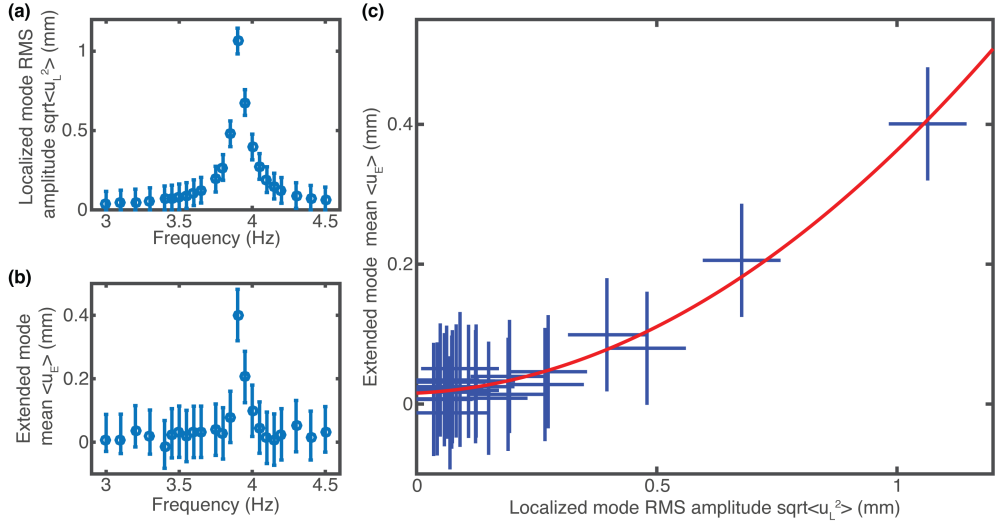

**FIG S5:** *Experimental determination of the nonlinear parameter $\gamma$.* **a** Frequency response of the localized mode. **b** Displacement of the extended mode as a function of the localized mode excitation frequency, measured simultaneously with panel **a**. **c** Experimental relationship between localized mode amplitude and extended mode static displacement (Crosses), and polynomial fit (Red line).